# Polaron Self-localization in White-light Emitting Hybrid Perovskites

Daniele Cortecchia,[1,2] Jun Yin,[3,4] Annalisa Bruno,[2] Shu-Zee Alencious Lo,[3] Gagik G. Gurzadyan,[3] Subodh Mhaisalkar,[2] Jean-Luc Brédas,[4] Cesare Soci[3,5,*]

[1] Interdisciplinary Graduate School, Nanyang Technological University, Singapore 639798

[2] Energy Research Institute @ NTU (ERI@N), Research Techno Plaza, Nanyang Technological University, 50 Nanyang Drive, Singapore 637553

[3] Division of Physics and Applied Physics, School of Physical and Mathematical Sciences, Nanyang Technological University, 21 Nanyang Link, Singapore 637371

[4] Laboratory for Computational and Theoretical Chemistry and Advanced Materials, Division of Physical Science and Engineering, King Abdullah University of Science and Technology, Thuwal 23955-6900, Saudi Arabia

[5] Centre for Disruptive Photonic Technologies, TPI, Nanyang Technological University, 21 Nanyang Link, Singapore 637371

*Corresponding Author: csoci@ntu.edu.sg

# Abstract

Two-dimensional (2D) perovskites with general formula $APbX_4$ are attracting increasing interest as solution processable, white-light emissive materials. Recent studies have shown that their broadband emission is related to the formation of intra-gap color centers; however, the nature and dynamics of the emissive species have remained elusive. Here we show that the broadband photoluminescence of the 2D perovskites $(EDBE)PbCl_4$ and $(EDBE)PbBr_4$ stems from the localization of small polarons within the lattice distortion field. Using a combination of spectroscopic techniques and *first-principles* calculations, we infer the formation of $Pb_2^{3+}$, $Pb^{3+}$, and $X_2^-$ (where X=Cl or Br) species confined within the inorganic perovskite framework. Due to strong Coulombic interactions, these species retain their original excitonic character and form self-trapped polaron-excitons acting as radiative color centers. These findings are expected to be applicable to a broad class of white-light emitting perovskites with large polaron relaxation energy.

# Introduction

After the demonstration of their outstanding performance in photovoltaic[1] and photodetector[2] devices, solution processable hybrid perovskites are currently in the spotlight for light-emitting applications,[3] such as light-emitting diodes (LEDs),[4,5] light-emitting transistors (LETs)[6] and tunable lasers.[7,8] Recent research suggests that two-dimensional (2D) perovskites,[9] in which the intrinsic layered configuration is responsible for strong exciton confinement within the inorganic interlayers,[10-12] have superior light-emitting properties than conventional three-dimensional perovskites, such as methylammonium lead iodide $MAPbI_3$.[13,14] Broadband, white-light generation has been reported in various 2D perovskites with general formula $APbX_4$ (X=Cl, Br and A=bidentate organic cation), which makes them particularly attractive for solid-state lighting and displays.[15] In earlier works on closely related layered materials with formula $Cd_{2-x}Zn_xE_2$(alkylamine) (E=S, Se, and Te),[16-18] the broadband luminescence was ascribed to deep surface states related to the large effective surface area intrinsic of the layered structure.[17,19] In the 2D perovskite (API)$PbBr_4$ (API=*N*-(3-aminopropyl)imidazole), the broadband photoluminescence spectrum was attributed by Li *et al*. to emission from the organic linker, upon energy transfer from the inorganic framework.[20] On the other hand, Dohner *et al*. related the highly efficient broadband emission of the perovskites (N-MEDA)[$PbBr_{4-x}Cl_x$] (N-MEDA=N1-methylethane-1,2-diammonium)[21] and (EDBE)$PbX_4$ (EDBE=2,2-(ethylenedioxy)bis(ethylammonium), X=Cl, Br)[22] either to distributed trap states or to the formation of self-trapped excitons. The hypothesis of exciton self-trapping was also supported by Yangui *et al.* who reported similar emission in $(C_6H_{11}NH_3)_2PbBr_4$,[23] and more recently by Hu *et al.* who probed photocarriers dynamics by THz spectroscopy.[24] These latest works have highlighted the importance of self-trapping in photoluminescence broadening; however, a thorough understanding of the nature of the emissive species and the localization sites is still lacking. In this work, we study the white-light emitting 2D perovskites (EDBE)$PbX_4$ (X=Cl,

Br) characterized by different orientations of the inorganic sheets, and provide further evidence for the photogeneration of polaronic species deriving from self-trapping of charge carriers at specific inorganic lattice sites. Details of the band structure and electronic transitions of these compounds are obtained by a combination of absorption and luminescence spectroscopy at various temperatures and *first-principles* density functional theory (DFT) calculations, which indicate that the optical properties are strongly correlated to charge-charge and charge-phonon interactions induced by the layered structure. Consequently, photoluminescence is generated efficiently upon excitation of the low energy excitonic absorption, whilst fast non-radiative thermalization dominates at higher energy excitation. Three main groups of emitters are found to contribute to the broad luminescence spectrum with large Stokes shifts; based on the charge density maps obtained from DFT, these are identified as $Pb_2^{3+}$, $Pb^{3+}$, and $X_2^-$ (where X=Cl or Br) small-polaron species confined within the inorganic perovskite framework.

## Results and discussion

The white light emitting hybrid perovskites (EDBE)PbCl$_4$ and (EDBE)PbBr$_4$ used in this study were synthetized by spin-coating to obtain high-quality and ultrasmooth thin films (see Supplementary Fig. 1). In agreement with previous results,[22] (EDBE)PbCl$_4$ crystallizes as a <100>-oriented perovskite,[10] forming a 2D structure of alternating flat inorganic layers and organic sheets of the di-ammonium cation $EDBE^{2+}$ (Fig. 1a, inset). Conversely, (EDBE)PbBr$_4$ has the typical structure of a <110>-oriented 2D perovskite,[10] characterized by rippled organic and inorganic sheets and distorted geometry (Fig. 1b, inset). The absorption spectra of both materials (Fig. 1a, b) can be qualitatively divided into two main regions: a high energy absorption continuum with an onset around 4 eV and 3.75 eV for the chlorine and bromine substituted compounds, and the sharp E-bands peaked at 3.72 eV and 3.32 eV, respectively. The E-bands are due to the excitonic absorption typical of layered perovskites; very much like in quantum well structures, excitonic absorption here is enhanced by quantum confinement effects due to the difference in optical gap and polarizability between the organic barriers and the inorganic wells.[25] The resulting *image charge effect*[11,25] substantially increases Coulomb interaction between electrons and holes within the wells, accounting for the large exciton binding energies (360 meV and 330 meV for (EDBE)PbCl$_4$ and (EDBE)PbBr$_4$, respectively) deduced from the low-temperature absorption spectra of the two compounds (see details provided in Supplementary Fig. 2). At low temperature, the excitonic bands split into three peaks equally spaced in energy (Fig. 1c, d), with separation comparable to the highest optical phonon energy of the inorganic precursors (~20-25 meV, see Raman spectra in Supplementary Fig. 1). This splitting is characteristic of a vibronic progression and indicates strong electron-phonon coupling induced by confinement within the inorganic wells. A similar splitting was also found in PbCl$_2$ and PbBr$_2$,[26] where charge self-trapping has been previously observed, and suggests a correlation with the corresponding inorganic precursors. The narrower spectral linewidth of the vibronic replicas in (EDBE)PbCl$_4$ is consistent with its ordered planar structure, compared to the corrugated structure of (EDBE)PbBr$_4$ (see Fig. 1, insets). An additional broad shoulder appears in this case at higher energy (3.48 eV, magenta line in Fig. 1d), but its origin is still unclear.

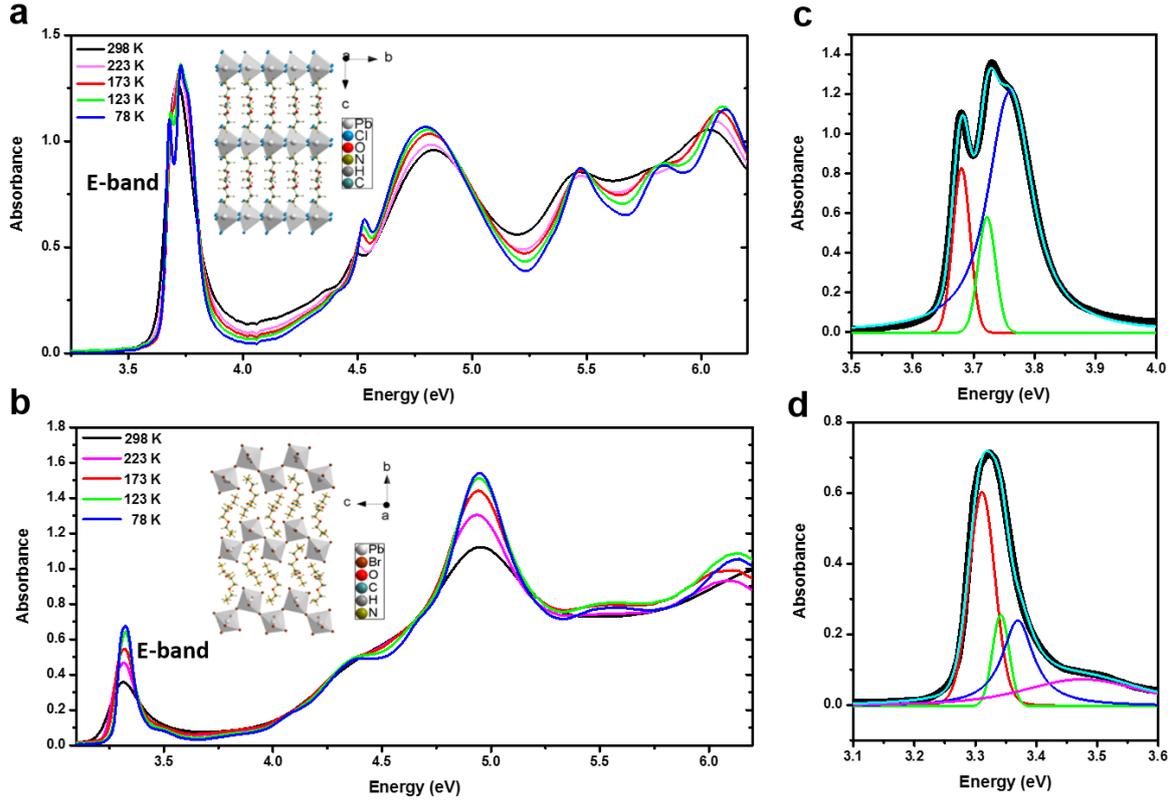

**Figure 1| Temperature dependent absorption spectra.** Steady state absorption and calculated oscillator strength of the optical transitions for **a**, (EDBE)PbCl$_4$ and **b**, (EDBE)PbBr$_4$. Chlorine replacement with bromine causes 400 meV redshift of the E-band, while it has minor effects on the high-energy part of the spectra. Insets show the layered structure of the <100>-oriented perovskite (EDBE)PbCl$_4$ (**a**, inset) and the <110>-oriented (EDBE)PbBr$_4$ (**b**, inset). Principal component fitting of the excitonic absorption at 78K showing vibronic splitting in 3 components (red, green and blue lines) for **c**, (EDBE)PbCl$_4$ and **d**, (EDBE)PbBr$_4$. Peaks are equally spaced by 40±2 meV and 29±2 meV in (EDBE)PbCl$_4$ and (EDBE)PbBr$_4$, respectively.

Calculations of the band structure and density of states at the density functional theory (DFT) level provide additional details on the higher-energy inter-band transitions. The unit cells of the optimized crystal structure of the two perovskites, starting point for the band structure calculations, are shown in Figs. 2a, b. When using the general gradient approximation (GGA)/Perdew-Burke-Ernzerhof (PBE) functional without spin-orbit coupling (SOC), (EDBE)PbCl$_4$ is calculated to have a direct bandgap (3.32 eV) at the Γ point, and (EDBE)PbBr$_4$ shows a direct bandgap (2.55 eV) at the A point of the respective Brillouin zones (Figs. 2c, d). Including SOC effects in the calculations retains the same band curvatures but with a change in conduction band degeneracies and reduced bandgaps of 2.49 eV for (EDBE)PbCl$_4$ and 2.00 eV for (EDBE)PbBr$_4$. The projected density of states (PDOS) reveals that in both materials the valence band consists of *p* orbitals of the halogen (Cl-3*p*, Br-4*p*) and Pb-6*s* orbitals while the conduction band is mainly composed of Pb-6*p* orbitals. This indicates that the high-energy absorption continuum is due to inter-band electronic transitions of the kind Pb$^{2+}$(6s)Cl$^-$(3p) → Pb$^{2+}$(6p) and Pb$^{2+}$(6s)Br$^-$(4p) → Pb$^{2+}$(6p) (similar to the situation in the parent lead halides PbX$_2$ compounds[27]), and confirms that the E-bands are due to excitons confined within the closely spaced inorganic layers. The organic EDBE layers contribute little to the top region of the valence band and the bottom region of the conduction band; this is similar to three-dimensional Pb-based hybrid perovskites, such as CH$_3$NH$_3$PbI$_3$.[28] However, even though the EDBE organic cations do not actively contribute to the inter-band transitions, which are primarily due to the inorganic scaffold, they are responsible for the emergence of the sharp E-band absorption, through the formation of highly confined excitonic states within the layered inorganic quantum wells of the 2D perovskite structure.

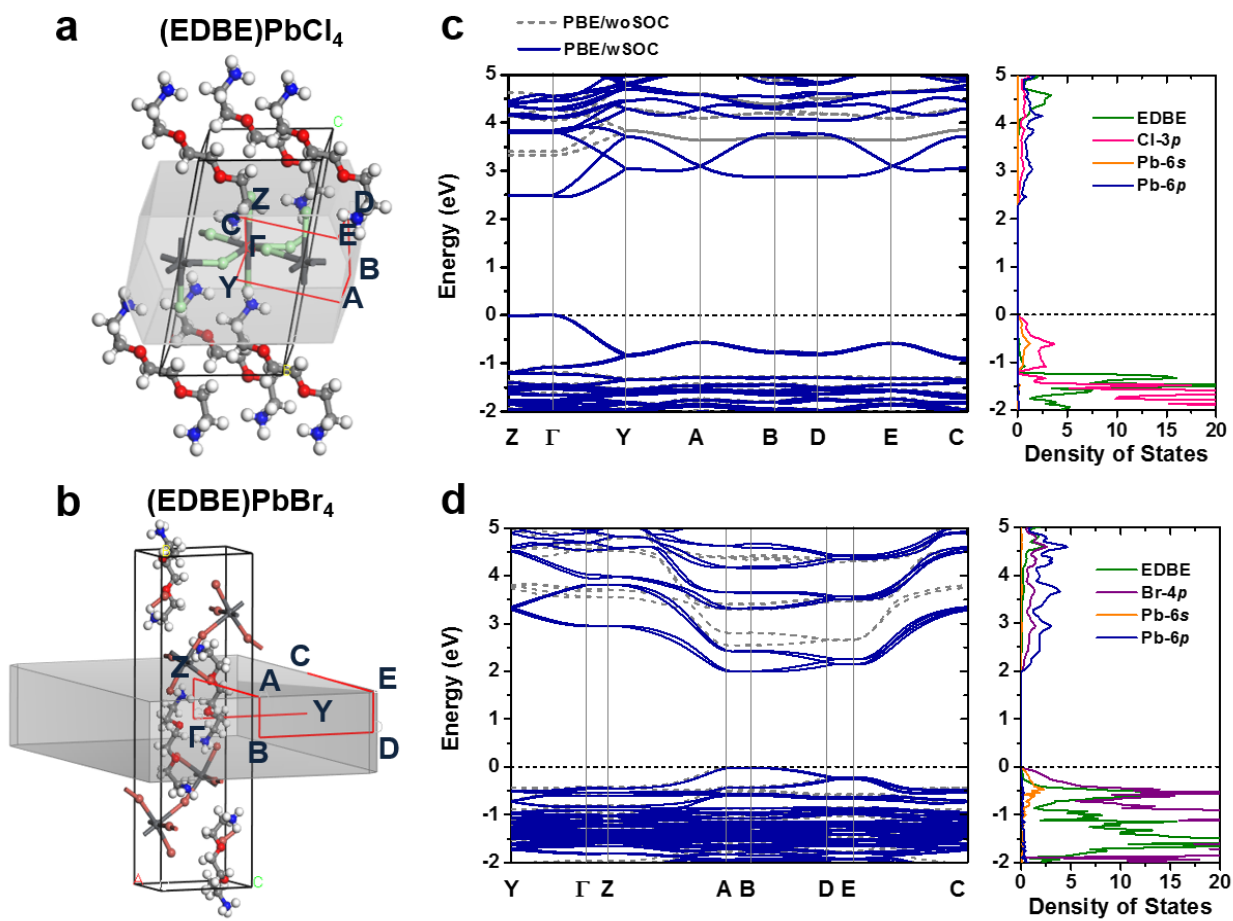

**Figure 2| Density functional theory (DFT) calculations at the GGA/PBE level.** Optimized crystal structure, obtained using the experimental cell parameters as starting point, for **a,** (EDBE)PbCl$_4$ and **b,** (EDBE)PbBr$_4$. Band structures (with and without inclusion of spin-orbital coupling – wSOC and woSOC) and projected density of states with SOC of **c,** (EDBE)PbCl$_4$ and **d,** (EDBE)PbBr$_4$. Even in the absence of SOC, the resulting bandgap values are lower than the experimental absorption edges, namely 3.32 eV for (EDBE)PbCl$_4$ and 2.55 eV for (EDBE)PbBr$_4$ (see Supplementary Fig. 2). PBE/wSOC lowers the bandgap energies to 2.49 and 2.00 eV for (EDBE)PbCl$_4$ and (EDBE)PbBr$_4$, respectively.

Upon photoexcitation, (EDBE)PbCl$_4$ and (EDBE)PbBr$_4$ emit extremely broadband luminescence throughout the entire visible range, with unusually large "Stokes shifts" of the emission peak relative to the maximum of the E-band (1.38 eV and 1.00 eV, respectively, see Fig. 3). This is in sharp contrast with the narrow emission spectra and small Stokes shift typical of 2D excitonic perovskites like phenethylammonium and butylammonium based perovskites.[29,30] Despite the broad absorption spectra shown in Fig. 1, efficient fluorescence is observed exclusively within a narrow band of excitation energies corresponding to the E-bands.

In particular, the photoexcitation maps of (EDBE)PbCl$_4$ (Fig. 3a,b) reveal two maxima peaking at excitation energies of 3.84 eV and 3.74 eV at low temperature, which trace the spectral shape of vibronic replicas in the excitonic absorption band (Fig. 1a). The slightly wider spacing of the two peaks in the excitation spectrum can be expected to be due to differences between the vibrational modes of the excited state compared to the ground state. Similarly, the photoexcitation map of (EDBE)PbBr$_4$ shows a single maximum at $E_{exc}$=3.32 eV (Fig. 3c,d), which matches perfectly the excitonic absorption lineshape (Fig. 1b). Unexpectedly, excitation at higher photon energies into the band continuum yields weaker and progressively narrower emission spectra. This is also reflected in time-resolved photoluminescence (TRPL) measurements, where efficient radiative recombination with characteristic decay time of $\tau \approx 3$ ns is achieved upon resonant excitation of the excitonic absorption, while excitation of higher energy transitions (e.g., $E_{exc}$=4.66 eV) results in extremely fast photoluminescence decay ($\tau \ll 40$ ps), likely due to deactivation through nonradiative decay channels (Supplementary Fig. 3 and Supplementary Table 1). This indicates that *the formation of radiative states is conditional to the creation of excitons*, alike conventional traps which would be populated upon thermalization from high-energy states, regardless of photoexcitation energy.

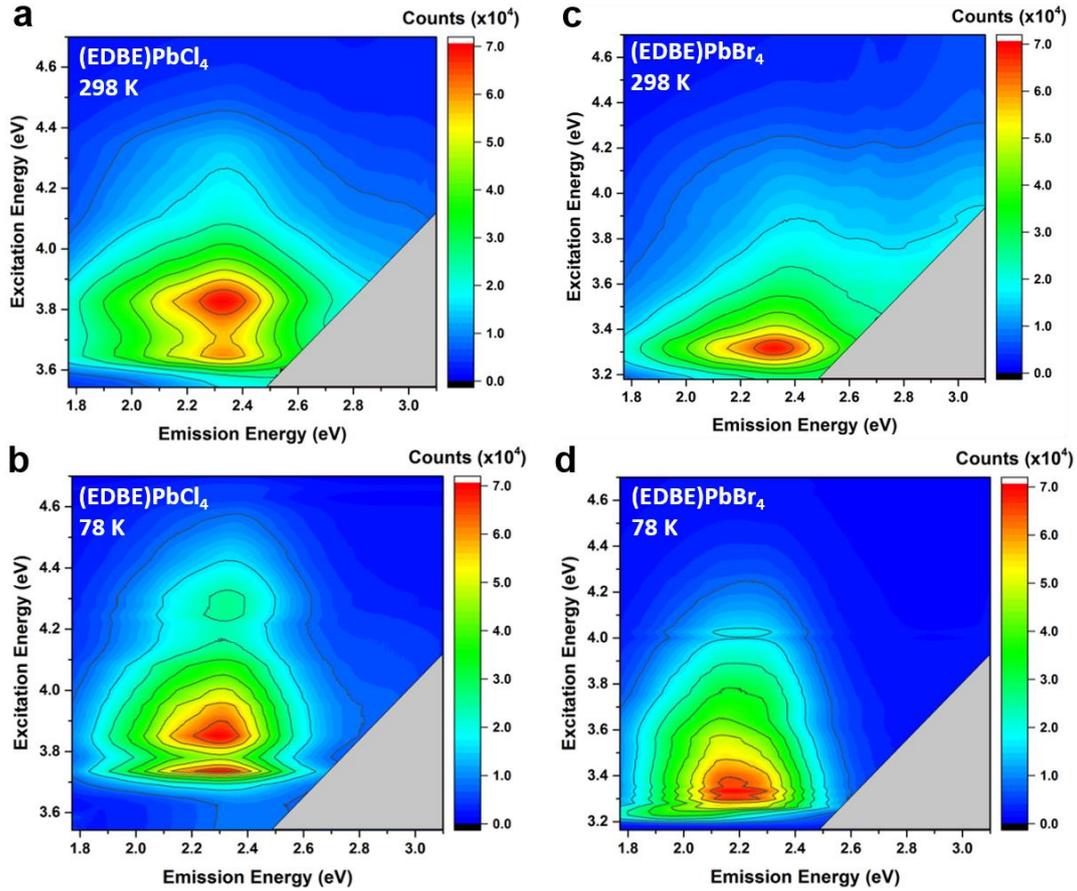

**Figure 3| Luminescence photoexcitation maps.** Contour plots of the photoluminescence emission intensity as a function of emission and excitation energies for (EDBE)PbCl$_4$ (left panels **a**, **b**) and (EDBE)PbBr$_4$ (right panels **c**, **d**). Maps in the top panels (**a**, **c**) were recorded at room temperature (T=298 K) while those in the bottom panels (**b**, **d**) were recorded at low temperature (T=78 K).

Notwithstanding the strong dependence of the energy of the E-band on material composition, the emission profiles of the two compounds are surprisingly similar (Figs. 4a and 4c and Supplementary Fig. 4). At room temperature, both emission spectra are peaked at 2.34 eV, with a full width at half maximum (FWHM) of 650 meV. This observation strongly suggests that the broadband radiative emission involves analogous intermediate states. To determine the emissive states, the luminescence spectra of (EDBE)PbCl$_4$ and (EDBE)PbBr$_4$ were analyzed by principal component fitting (Supplementary Figs. 5 and 6). At each temperature, three main components contributing to white-light emission can be identified, hereafter denoted W1, W2, and W3 (the central energy of these components in (EDBE)PbBr$_4$ is indicated by dashed lines in Fig. 4a). The large and irregular energy spacings between W1, W2, and W3 (~160-190 meV) rule out their possible vibronic nature, as seen earlier in the case of E-band absorption (Fig. 1b, d). As expected for deep levels, the spectral position of W1, W2 and W3 are nearly independent of temperature (Supplementary Fig. 5 and 6; note that the change in relative weight of each components is responsible for the apparent red-shift of the total emission at lower temperatures). On the other hand, the total photoluminescence intensity increases significantly at low temperature due to the reduction of line broadening and non-radiative recombination. Fitting of the Arrhenius plots[31] of the integrated photoluminescence intensity yields thermal activation energies of $E_a$=147 meV and $E_a$=105 meV for (EDBE)PbCl$_4$ and (EDBE)PbBr$_4$, respectively (Supplementary Fig. 4). Such large activation energies imply the existence of radiative states deep in the gap, with energy much larger than thermal vibrations.

The existence of multiple radiative components was further confirmed by the spectral dependence of photoluminescence decay dynamics. The transient photoluminescence decay of (EDBE)PbBr$_4$ (Fig. 4b) was found to contain at least three components, with characteristic decay times of $\tau_1$=40 ps, $\tau_2$=0.74 ns, and $\tau_3$=3.24 ns, as determined by global fitting of the entire dataset in Fig. 4b with multiple exponential waveforms (Table 1). We ascribe the ultrafast decay, with larger weight at high emission energies (2.82 eV and 2.64 eV), to hot exciton emission, which would have marginal contribution to steady-state luminescence. Conversely, the slower components, which prevail at low emission energies, can be related to the W1-W3 emitters. Since the spectral dependence of amplitude A3 resembles closely that of the W3 emission profile (Table 1 and Supplementary Fig. 6), we assign $\tau_3$ to this particular emitter, while $\tau_2$ could result from the superposition of comparable characteristic lifetimes of W1 and W2, which cannot be fully resolved by the fitting.

Transient absorption (TA) measurements performed with excitation resonant to the E-bands show the formation of an unstructured excited-state absorption spanning the entire visible spectral range (Fig. 4c and Supplementary Fig. 7), consistent with the formation of color centers distributed throughout the bandgap. The absence of stimulated emission further confirms that the broadband photoluminescence is not directly correlated to singlet excited states. The formation time of the trapped states, as determined from the rise of the photoinduced absorption, is comparable to our instrumental resolution (~100 fs). This is in agreement with the ultrafast self-trapping of charge carriers reported by Hu *et al*,[24] which was attributed to the coupling to vibrational modes of the inorganic scaffold. The TA decay dynamics contain two components, and are nearly independent of probe photon energy (Fig. 4d and Supplementary Table 2). In the case of (EDBE)PbBr$_4$, the fast component ($\tau_1 \approx 10$ ps) can be related to the excitonic emission ($\tau << 40$ ps) observed in TRPL with non-resonant excitation. We tentatively ascribe it to the population of the first excited singlet state (S$_1$) undergoing absorption to higher singlet states (S$_N$). Conversely, the slower process with decay time $\tau_2 = 0.74$ ns matches well the $\tau_2$ decay time measured in TRPL, and may be indicative of the excited-state absorption from the intra-gap trap states.

**Table 1| Time resolved photoluminescence (TRPL) parameters for (EDBE)PbBr$_4$.** The characteristic lifetimes ($\tau$) and amplitudes (A) were extracted from the global fitting of the five decays with a three-exponential decay function ($E_{exc}$ = 3.26 eV). Fit result for $\tau_1$=40 ps indicates that in fact $\tau_1$<<40 ps, since $\tau_1$ is very close to the time resolution of our setup (IRF=20-30 ps).

| Emission Energy (eV) | $\tau_1$ (ns) | $A_1$ | $\tau_2$ (ns) | $A_2$ | $\tau_3$ (ns) | $A_3$ |
|---|---|---|---|---|---|---|
| **2.82** | 0.04 | 0.81 | 0.74 | 0.10 | 3.24 | 0.10 |
| **2.64** | 0.04 | 0.49 | 0.74 | 0.28 | 3.24 | 0.23 |
| **2.34** | 0.04 | 0.00 | 0.74 | 0.09 | 3.24 | 0.91 |
| **2.25** | 0.04 | 0.20 | 0.74 | 0.18 | 3.24 | 0.63 |
| **2.14** | 0.04 | 0.00 | 0.74 | 0.41 | 3.24 | 0.59 |

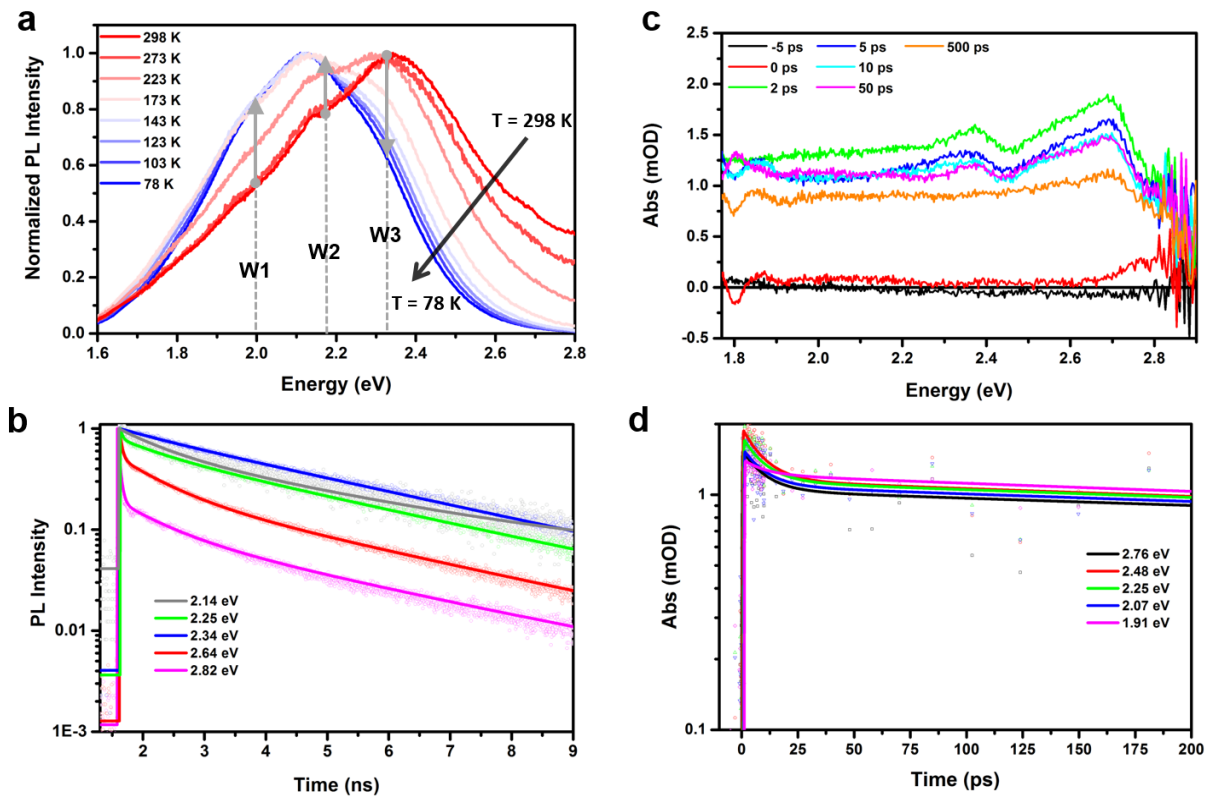

**Figure 4| Multicomponent analysis of broadband photoluminescence and transient absorption (TA) measurements of (EDBE)PbBr$_4$. a,** Temperature-dependent steady state photoluminescence spectrum. The dashed lines indicate the temperature-independent energy of the three principal components (W1=1.98 eV, W2=2.14 eV and W3=2.32 eV) determined from the analysis of all photoluminescence spectra (Supplementary Fig. 6). **b,** spectral dependence of time resolved photoluminescence (TRPL) performed under excitation energy $E_{exc}$=3.26 eV. **c,** transient absorption (TA) spectrum under resonant excitation of the excitonic peak ($E_{exc}$=3.26 eV) and **d,** corresponding decay dynamics at different probing energies.

The nature of the emissive states is further elucidated via our *first-principles* DFT-PBE calculations of the charge density distributions. In a way similar to organic semiconductors,[32,33] polaronic effects have been recently considered in order to understand the charge transport and light emission characteristics of hybrid organic-inorganic perovskites.[24,34-40] Notably, it was proposed that the formation of small polarons stabilized by collective rotations of methylammonium cations[40] may play a role in the photodegradation of MAPbI$_3$.[39] In crystals where photoexcitations cause significant lattice deformation, holes and electrons may be localized at specific lattice sites by their own distortion field, giving rise to self-trapped electrons (STEL) and self-trapped holes (STH).[41,42] This process can also bring temporary short-range chemical bonding between nearest neighbor ions (molecular polaron).[43] When self-localized carriers are bound electron-hole pairs, the resulting species are described as polaron-excitons, PE. Self-trapping phenomena have been extensively studied and frequently observed in alkali, alkaline-earth, and perovskite-structure halides (e.g., KCl, CaF$_2$, KMgF$_3$).[42,44] Similarly, charge self-trapping at low temperature in lead halides PbX$_2$ (X=Cl, Br) is known to yield large Stokes shift and broad photoluminescence.[26,45-47] In this case, electron-spin resonance (ESR) measurements have also identified the self-trapping centers to be Pb$_2^{3+}$ for STEL, and Pb$^{3+}$ and X$_2^-$ (X=Cl, Br) for STH.[42,45,47-52] The latter, also known as V$_k$ centers,[53,54] are commonly observed in alkali halides,[44] where hole trapping strengthens the interaction between halide pairs leading to the formation of dimer species X$_2^-$ within the ionic crystal.[42]

We discussed earlier how the organic cations in 2D EDBE perovskites are responsible for the formation of a layered structure, where charges are strongly confined in inter-layer potential wells and subject to strong electron-phonon interactions. Yet, optical transitions are mainly related to the inorganic layers originating from the lead halide precursors PbX$_2$, and the photoexcitation properties are very similar to those of lead halides PbX$_2$.[55] It is then reasonable to expect the occurrence of charge self-trapping effects in EDBE perovskites, similar to those in PbX$_2$. To prove this hypothesis, we computed the charge density maps for holes and electrons in (EDBE)PbCl$_4$ (Fig. 5a-c) and (EDBE)PbBr$_4$ (Fig. 5d-f) in which dimerization has been introduced as a perturbation in the crystal lattice. We first constructed the 2D perovskite supercell models and exerted local perturbations, where selective bond lengths of nearest neighbor atoms (Pb-Pb, Pb-X, and X-X) were shortened to induce the local structural deformations; then, the charge density distributions of the VBM and CBM were plotted based on the PBE/wSOC results.

Compared to the unperturbed systems, where charges are fully delocalized in the inorganic framework (Supplementary Fig. 8 and 9), the Pb-Pb dimerization causes selective localization of electron density close to the strained lattice point in both perovskites, which is consistent with the self-trapping of electrons at Pb$_2^{3+}$ sites (Fig. 1a,d) previously observed in lead halides.[48-50] On the other hand, hole self-localization occurs with Cl-Cl and Br-Br pairing, driving the formation of V$_k$ centers Cl$_2^-$ and Br$_2^-$ for the chloride and bromide perovskite, respectively (Fig. 5b,e). Hole density localization can also occur at Pb$^{2+}$ sites leading to the formation of Pb$^{3+}$ centers, coupled with the lattice deformation involving shortening of Pb-X bond lengths around the localization site (Fig. 5c,f). A similar behavior has been previously observed in AgCl, where hole localization at Ag$^+$ sites causes the tetragonal distortion of the surrounding lattice.[56] Overall, X$_2^-$ and Pb$^{3+}$ correspond to two distinct self-trapped holes. Interestingly, the results of the calculations in Fig. 5a-f show that similar polaronic species form in both EDBE perovskites, irrespective of the <100> or <110> orientations, which indicates that exciton localization occurs independently of the planarity of the octahedra coordination planes (Fig. 1).[30]

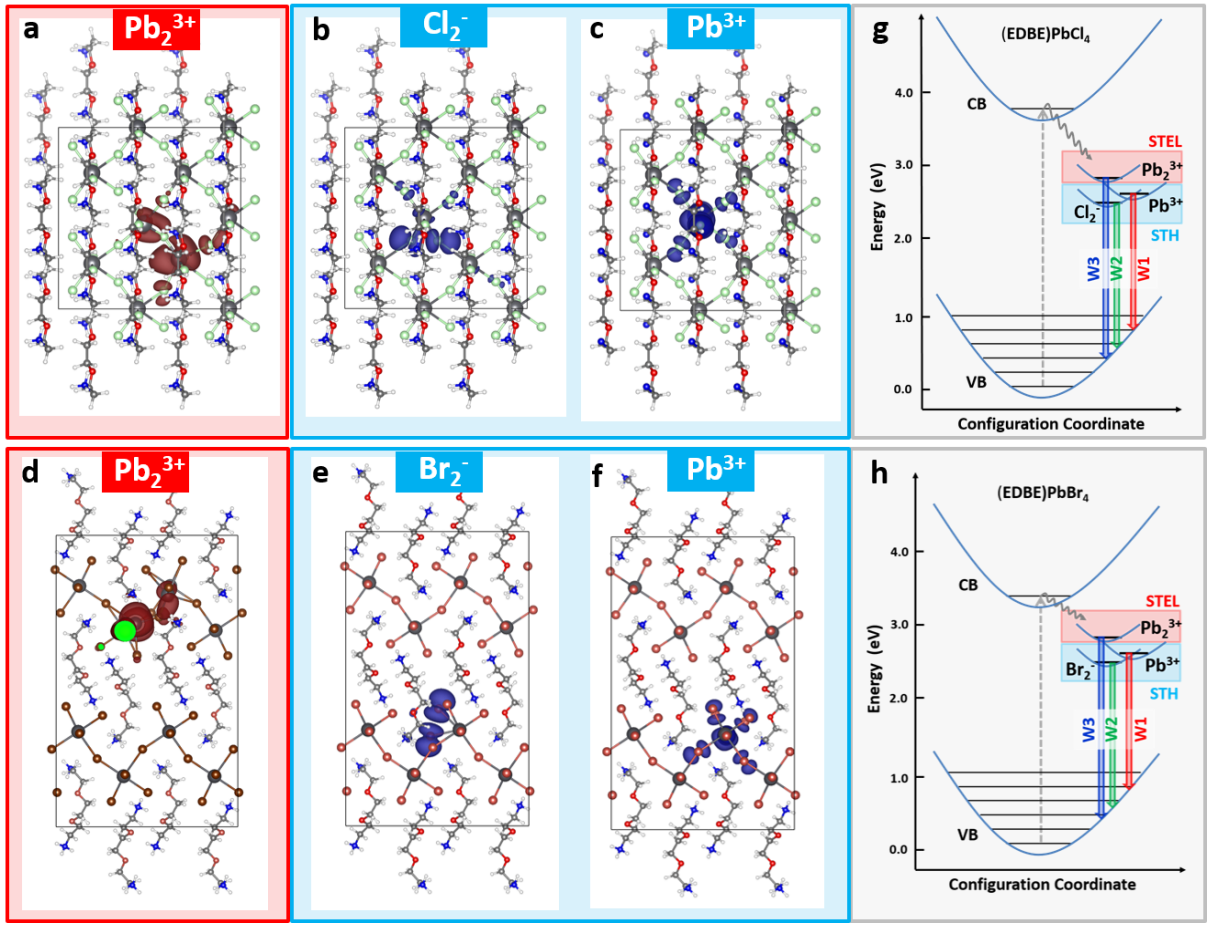

**Figure 5| Charge density mappings and exciton relaxation model.** Charge density calculated for (EDBE)PbCl$_4$ (**a**, electrons, **b, c** holes) and (EDBE)PbBr$_4$ (**d**, electrons, **e, f** holes) upon perturbation of selected crystal lattice sites at the PBE/wSOC level. The applied perturbations involve shortening of the Pb-Pb (**a, d**), X-X (**b, e**) and Pb-X (**c, f**) distances, with the resulting charge localization showing the formation of self-trapped electrons (STEL) Pb$_2^{3+}$ and self-trapped holes (STH) X$_2^-$, Pb$^{3+}$ in the form of small polarons localized in the metal-halide framework. Schematic representation of the emissive process via polaron formation involved in the white-light generation of **g**, (EDBE)PbCl$_4$ and **h**, (EDBE)PbBr$_4$. Conversion of Wannier-Mott exciton into polaronic exciton (PE) proceeds via formation of self-trapped electrons (STEL) on Pb$_2^{3+}$ sites, and self-trapped holes (STH) on Pb$^{3+}$ and X$_2^-$ (X = Cl, Br) states. The relative energies of these species were also determined (Supplementary Fig. 9). Radiative decay from each of these self-trapped states results in the three main emission bands observed in steady-state and time-resolved photoluminescence (W1, W2, and W3).

At this stage, by combining our experimental and computational results, we can propose the following model for the mechanism of white-light emission in (EDBE)PbX$_4$ (Fig. 6g, h). Upon resonant photoexcitation of the E-band, Wannier-Mott excitons are generated within the potential wells of the inorganic interlayers of 2D perovskites. Electrons and holes in the bound exciton pairs self-localize in the crystal lattice, converting these Wannier-Mott excitons into polaron-excitons with reduced average binding energy of the order of 100-150 meV (corresponding to the activation energy determined in Supplementary Fig. 4). This leads to the formation of self-trapped electron (STEL) $Pb_2^{3+}$ and self-trapped hole (STH) $Pb^{3+}$ and $X_2^-$ states with relative energies $E_{X_2^-}^{STH} < E_{Pb^{3+}}^{STH} < E_{Pb_2^{3+}}^{STEL}$, which define the specific polaronic emissive states (Supplementary Fig. 10). It is likely that inter-band absorption at higher photon energies activates alternative non-radiative relaxation channels, which prevent the formation of self-trapped polaron-excitons. Electron-hole annihilation associated to STEL contributes to light emission corresponding to the W3 band; similarly, the polaronic excitons deriving from STH states recombine radiatively leading to the W1 and W2 photoluminescence bands.[57] It appears that at low temperature hole trapping is more efficient than electron trapping, increasing the weight of STH luminescence; this is in agreement with previous calculations predicting smaller localization energy of the holes compared to the electrons in alkali halides.[58] Strong electron-phonon coupling involved in polaron formation explains the large Stokes shift observed experimentally, and further broadens the bandwidth of radiative transitions. Although the organic layer is not actively involved in polaron formation, charge confinement in the layered structure significantly increases the carrier effective masses,[59] strengthening electron-phonon coupling and fostering polaron formation even at room temperature. Thus, this model is not expected to be limited to the EDBE-based perovskites of this study, but to apply to a wide family of perovskites with large electron-phonon coupling, including 2D white-light emitting perovskites such as (API)$PbBr_4$, (N-MEDA)[$PbBr_{4-x}Cl_x$], and ($C_6H_{11}NH_3$)$_2PbBr_4$, and 3D perovskites in which the lattice is distorted by photoexcitation or other physical (e.g., hydrostatic pressure) or chemical (e.g., compositional or stoichiometric tuning) means.

# Conclusion

We have conducted a combined, systematic spectroscopic and computational study of the white-light emission properties of layered organic-inorganic perovskites (EDBE)PbCl$_4$ and (EDBE)PbBr$_4$. The results allow us to formulate a comprehensive model of exciton relaxation dynamics leading to their unusually large Stokes shift and white broadband photoluminescence. We provide both theoretical and experimental evidence for the formation of self-localized polarons and identify polaron-excitons Pb$_2^{3+}$, Pb$^{3+}$, and X$_2^-$ (X = Cl, Br) localized within the inorganic lattice as the intra-gap emissive species.

Overall, these findings prompt for wider consideration of the role of electron-phonon interactions and structural deformations on the optoelectronic properties of perovskites. Structural distortion of the soft lattice may be engineered to improve charge transport properties of highly luminescent 2D perovskites for light-emitting diodes and displays. This concept may also be explored to tune the ratio of excitons and charge carriers in 3D perovskite materials, with potential relevance to transistors and solar cell devices.

# Methods

**Perovskite synthesis.** 2,2'-(ethylenedioxy)bis(ethylamine) (98%), hydrochloric acid (HCl, 37% H$_2$O) hydrobromic acid (HBr, 48% in H$_2$O), lead(II) chloride (PbCl$_2$, 99.999% trace metal basis), lead(II) bromide (PbBr$_2$, 99.999% trace metal basis) and dimethyl sulfoxide (DMSO, anhydrous 99,9%) were purchased from Sigma-Aldrich. Quartz cuvettes (four sides clear, 20mm path length, wavelength range: 170-2700nm) for optical measurements in N$_2$ inert atmosphere were purchased by Achema Pte Ltd, and quartz substrates from Crystran Ltd. (EDBE)Cl$_2$ was synthetized by reaction of 1 equivalent of HCl (37% in H$_2$O) with 2,2'-(ethylenedioxy)bis(ethylamine) for 2h at 0 °C. (EDBE)Br$_2$ was synthetized by reaction of 3 equivalents of HCl (48% in H$_2$O) with of 2,2'-(ethylenedioxy)bis(ethylamine) for 2h at 0 °C.

The resulting white hygroscopic salts were collected with a rotary evaporator, dried in vacuum oven overnight at 60 °C and stored in glove box ($N_2$ atmosphere). (EDBE)PbCl$_4$ was prepared by mixing (EDBE)Cl$_2$ and PbCl$_2$ (1:1 molar ratio), while (EDBE)Br$_2$ and PbBr$_2$ (1:1 molar ratio) were mixed to prepare (EDBE)PbBr$_4$. In all cases, solutions with the desired concentration (0.1M, 0.25M and 0.5M) were prepared in DMSO dissolving the powders at 100 °C for 1h. The films were prepared by spinning the hot solution (100 °C) on cold substrates at 4000 rpm, 60s, and annealed for 15 minutes on the hotplate (100 °C). The perovskite deposition was performed in glove-box under $N_2$ environment.

**Structural and morphological characterization.** X-ray diffraction characterization of perovskite thin films was performed using a BRUKER D8 ADVANCE with Bragg-Brentano geometry employing Cu K$\alpha$ radiation ($\lambda$=1.54056 Å), step increment of 0.02° and 1 s of acquisition time. Raman spectra were obtained in a Renishaw Raman microscope configured with a charge coupled device array detector. A laser line ($\lambda_{ex}$=532 nm) was used for excitation with power below 1 mW. Raman signals were collected by a Leica 1003 objective lens (NA=50.85) and dispersed by 2400 line/mm gratings with frequency resolution of 0.8 cm$^{-1}$. The integration time was 20 s.

**Optical characterization.** Thin film samples were deposited on quartz substrates and mounted into a liquid nitrogen-cooled Linkam Stage (FTIR 600) that allow to reach operating temperatures down to 77 K. Absorption spectra were recorded by an UV-VIS-NIR spectrophotometer (UV3600, Shimadzu) using a scanning resolution of 0.1 nm. Steady-state photoluminescence spectra were recorded by a Fluorolog-3, (HORIBA Jobin Yvon) spectrofluorometer with wavelength resolution 0.5 nm. Principal component fitting of absorption and emission lineshapes was performed using Voigt line profiles to account for the convolution of multiple broadening mechanisms. Time-resolved fluorescence was measured by time-correlated single photon counting (TCSPC) technique with resolution of 30 ps (PicoQuant PicoHarp 300). The second harmonic of Titanium sapphire laser (Chameleon, Coherent Inc.) at 400 nm (100 fs, 80 MHz) was used as the excitation source. The kinetics of fluorescence from 438 to 579 nm were recorded. Deconvolution/fit procedure was applied in order to obtain time components of fluorescence decay. A home built pump-probe setup is used for our transient absorption measurement. A commercial amplifier system, Quantronix Integra-C is used as the laser source at the repetition rate of 1 KHz and pulse width of 100 fs. Broadband white light with highest cutoff photon energy at 3.54 eV (350 nm) is generated in a 3mm thick Calcium Fluoride crystal via white light continuum generation. To prevent laser induced damaged, the crystal is constantly spinning during the duration of measurement. A commercial spectrometer, Horiba CP140-104 equipped with a silicon photodiode array is used to record the transient absorption spectra. To remove higher diffraction order artefact in the measurement, a shortwave pass filter is used to cut off the low energy end of the white light spectrum at 1.77 eV (700 nm). Pump wavelength of 266 nm is generated by first generating 400 nm using a type I BBO crystal and subsequently followed with third harmonic generation using another BBO crystal cut for sum frequency generation with the generated 400 and residue 800 nm. Pump wavelength of 330 and 370 nm was generated by doubling the output of a commercial optical parametric amplifier, Quantronix Palitra running at 660 nm and 740 nm respectively with a 1mm thick BBO crystal cut at 29.2°. To reduce white light stability artefact in the measured transient absorption spectra, selected spectra at different time delay was obtained at 3000 measurement cycles while the kinetic spectra was obtained at 900 measurement cycles for reasonable measurement time.

**Computational methods.** The crystal structural optimization, band structure, and charge density calculations were performed using the Vienna *ab initio* Simulation Package (VASP).[60,61] The projector augmented-wave (PAW) method was used with PBE exchange-correlation functional to describe the electron-ion interactions with electronic orbitals of H (1s); O, N and C (2s, 2p); Cl (3s, 3p); Br (4s, 4p) and Pb (5d, 6s, 6p). The plane-wave basis set cutoffs of the wave functions were set to 500 eV; 4×4×4 and 4×2×4 Monkhorst–Pack grids were chosen for sampling the Brillouin zone of (EDBE)PbCl$_4$ and (EDBE)PbBr$_4$, respectively. The experimental crystal structures for both monoclinic crystals at room temperature were used as an initial guess. The atomic relaxation calculations were performed until the residual atomic forces were less than 0.01 eV/Å. The hole and electron densities under four possible perturbations (corresponding to shortenings of specific bond lengths) were plotted to simulate the formation of self-trapped charges. The molecular graphics viewer VESTA was used to plot the molecular structures and charge densities.

## Acknowledgements

Research at NTU was supported by the Ministry of Education (ref. Nos. MOE2013-T2-1-044 and MOE2011-T3-1-005) and the National Research Foundation (ref. No. NRF-CRP14-2014-03) of Singapore. JY and JLB acknowledge support from King Abdullah University of Science and Technology and thank the IT Research Computing Team and Supercomputing Laboratory at KAUST for computational and storage resources.

## Author contributions

CS, AB and DC designed the experiments. DC synthesized the perovskite precursors, prepared and characterized the films. AB and DC collected and analyzed absorption and steady-state photoluminescence measurements. JY, JB performed DFT modeling and calculations. GG conducted time-resolved photoluminescence measurements and SL performed transient absorption measurements. DC, AB and CS analyzed the data and drafted the manuscript. All the authors contributed to interpretation of the results and revision of the manuscript. CS supervised the work.

# Supporting Information

# Polaron Self-localization in White-light Emitting Hybrid Perovskites

Daniele Cortecchia,[1,2] Jun Yin,[3,4] Annalisa Bruno,[2] Shu-Zee Alencious Lo,[3] Gagik G. Gurzadyan,[3] Subodh Mhaisalkar,[2] Jean-Luc Brédas,[4] Cesare Soci[3,5,*]

[1] *Interdisciplinary Graduate School, Nanyang Technological University, Singapore 639798*

[2] *Energy Research Institute @ NTU (ERI@N), Research Techno Plaza, Nanyang Technological University, 50 Nanyang Drive, Singapore 637553*

[3] *Division of Physics and Applied Physics, School of Physical and Mathematical Sciences, Nanyang Technological University, 21 Nanyang Link, Singapore 637371*

[4] *Laboratory for Computational and Theoretical Chemistry and Advanced Materials, Division of Physical Science and Engineering, King Abdullah University of Science and Technology, Thuwal 23955-6900, Saudi Arabia*

[5] *Centre for Disruptive Photonic Technologies, TPI, Nanyang Technological University, 21 Nanyang Link, Singapore 637371*

[*]Corresponding Author: csoci@ntu.edu.sg

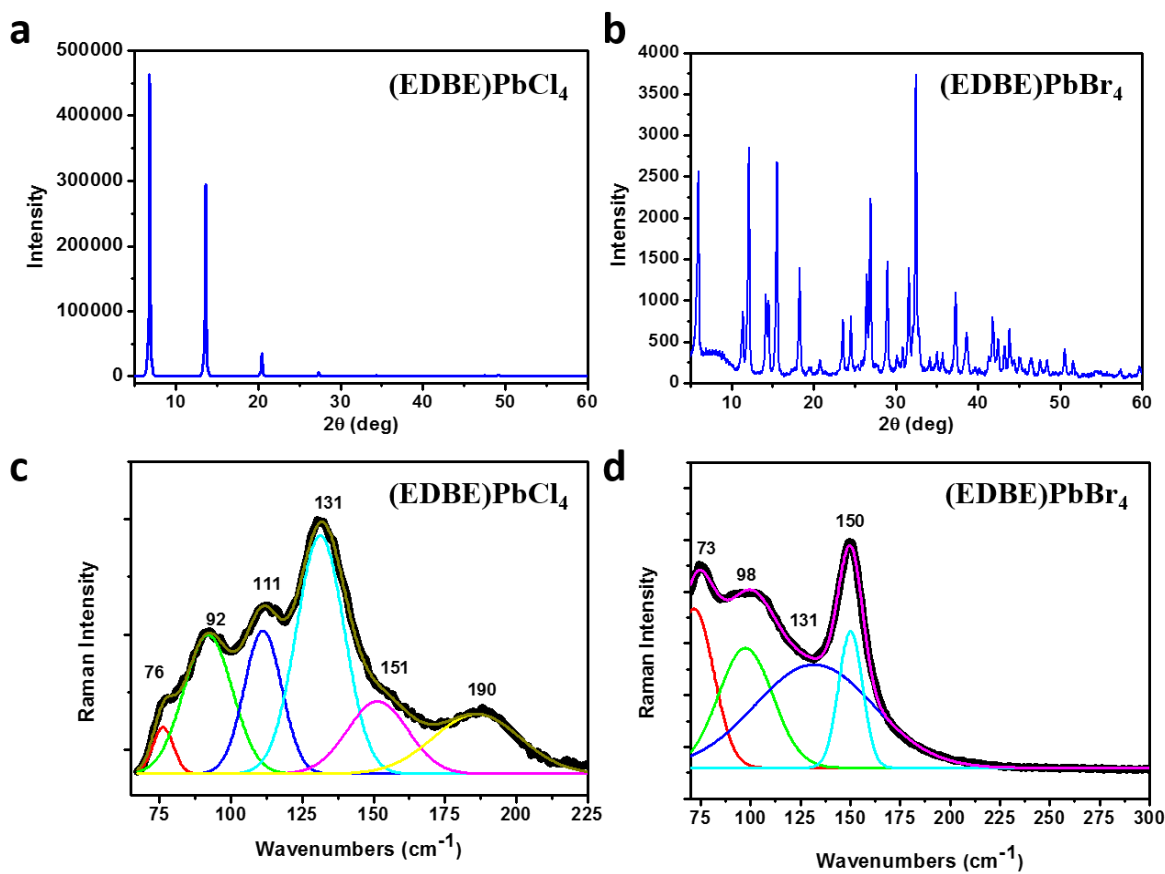

**Figure S1| Material characterization.** XRD patterns of drop casted films of **a,** (EDBE)PbCl$_4$ and **b,** (EDBE)PbBr$_4$ and correspondent Raman spectra (**c** and **d**).

X-ray analysis on drop-casted films confirmed the achievement of the desired phase (EDBE)PbCl$_4$ and (EDBE)PbBr$_4$ in agreement with the previous report on these materials. (EDBE)PbCl$_4$ (**a**), crystallizes as <100>-oriented perovskite with monoclinic crystal system, space group *C*2 and lattice parameters a = 7.748 Å, b = 7.523 Å, c = 13.346 Å, β = 102.68°. The drop casted film shows a strong preferential orientation toward the *00l* direction indicating a perfect alignment of the organic and inorganic sheet parallel to the substrate. (EDBE)PbBr$_4$ crystallizes as <110>-oriented perovskites (**b**) with monoclinic crystal system, space group *P*2$_1$/*c*, and lattice parameters a = 6.142 Å, b = 28.906 Å, c = 8.701 Å, β = 91.69° and does not show any strong preferential orientation in the drop-casted film.

The Raman spectra show the Pb-Cl and Pb-Br stretching indicating the formation of the inorganic framework: (EDBE)PbCl$_4$ spectrum shows peaks at 93, 112, 130 and 190 cm$^{-1}$ (**c**), while (EDBE)PbBr$_4$ has peaks at 74, 99 and 150 cm$^{-1}$ (**d**).

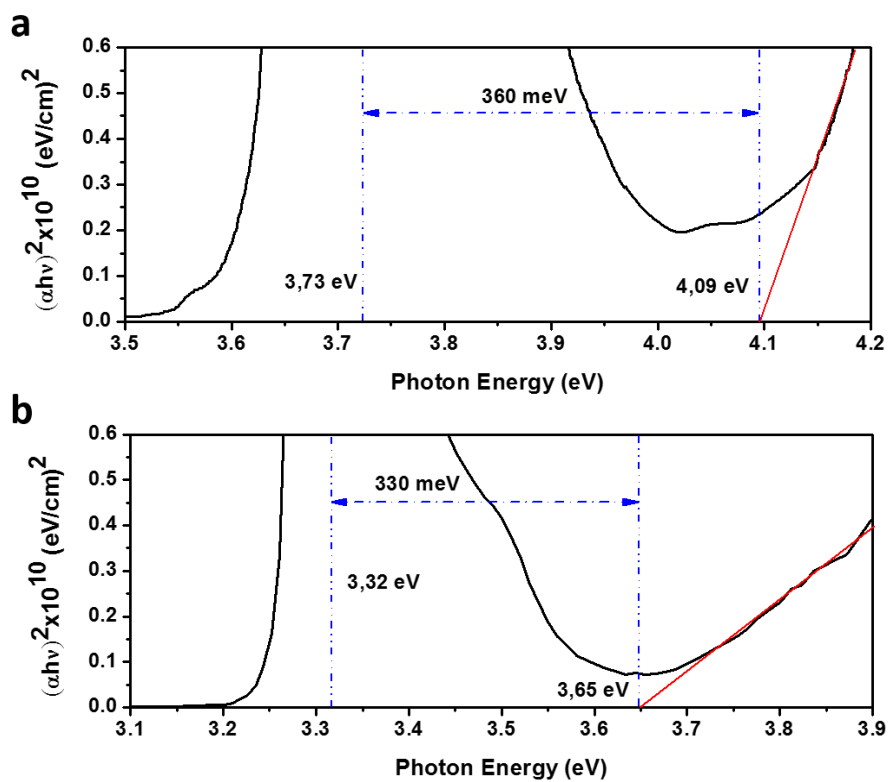

**Figure S2| Exciton binding energy determination.** Tauc Plots construction and exciton binding energy (EBE) at 78 K for **a,** (EDBE)PbCl$_4$ and **b,** (EDBE)PbBr$_4$. The exciton binding energy was estimated taking the difference between the peak position of the excitonic absorption (E-band) and the onset of the high energy absorption continuum. The extracted EBE is 360 meV and 330 meV for (EDBE)PbCl$_4$ and (EDBE)PbBr$_4$, respectively.

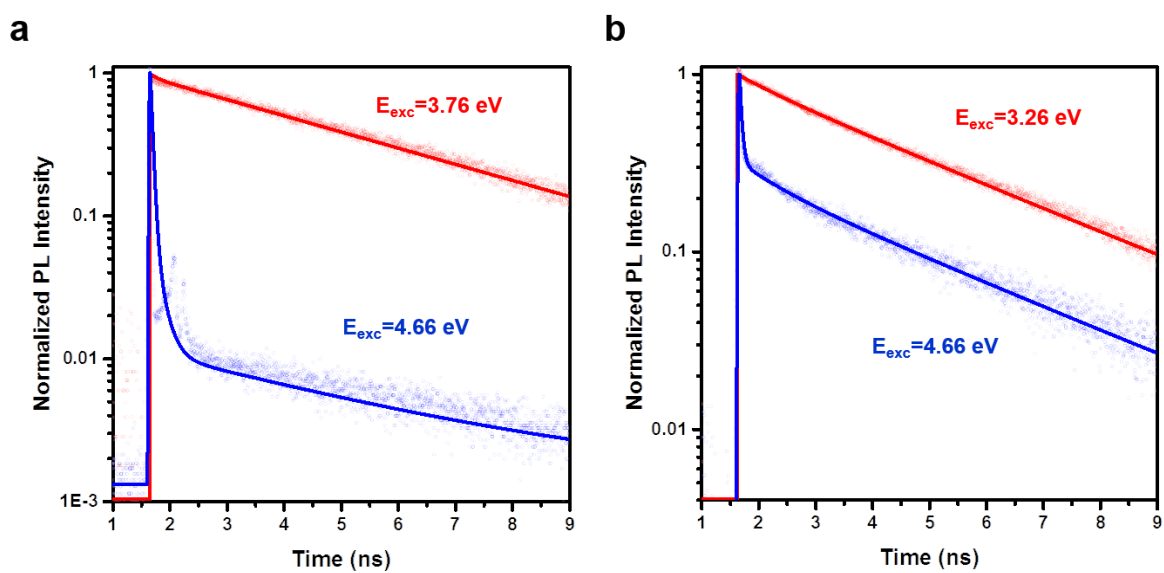

**Figure S3| Ultrafast dynamics dependence on excitation energy. a,** (EDBE)PbCl$_4$ fluorescence decays after excitation at 4.66 eV (blue) and excitation at 3.76 eV (red) with detection at 2.34 eV. **b,** (EDBE)PbBr$_4$ fluorescence decays upon excitation at 4.66 eV (blue) and 3.26 eV (red) with detection at 2.34 eV.

**Table S1| Time resolved photoluminescence (TRPL) parameters for different excitation energy.** The characteristic lifetimes (τ) and amplitudes (A) were extracted from the fitting of the dynamics at the probing energy of 2.34 eV.

| Material | Excitation Energy (eV) | τ 1 (ns) | A1 | τ 2 (ns) | A2 | τ 3 (ns) | A3 |
|---|---|---|---|---|---|---|---|
| (EDBE)PbCl$_4$ | 4.66 | 0.04 | 0.93 | 0.15 | 0.06 | 3.8 | 0.01 |
|  | 3.76 | 0.04 | 0 | 0.15 | 0.06 | 3.8 | 0.94 |
| (EDBE)PbBr$_4$ | 4.66 | 0.04 | 0.79 | 0.74 | 0.04 | 3.2 | 0.17 |
|  | 3.26 | 0.04 | 0 | 0.74 | 0.09 | 3.2 | 0.91 |

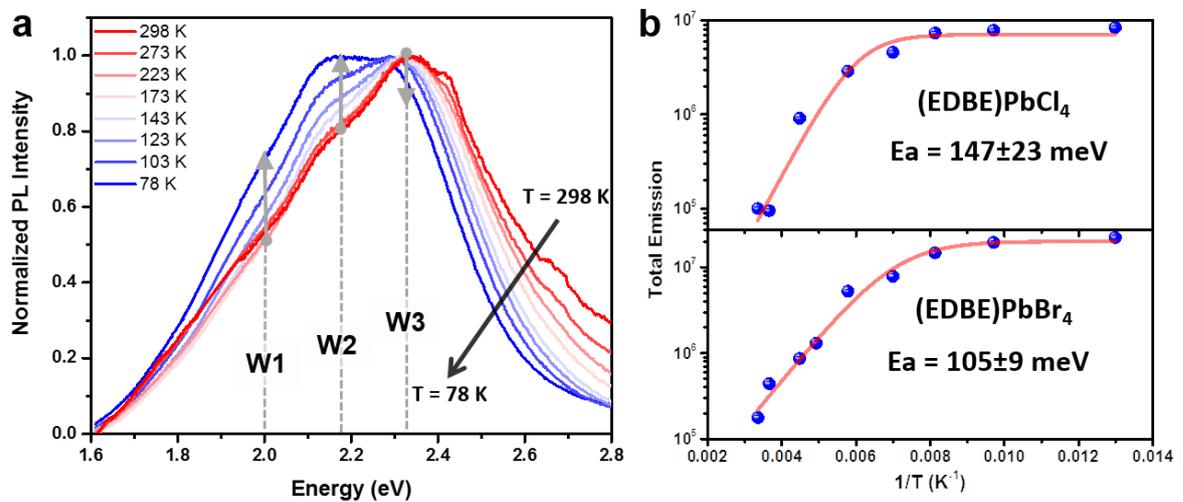

**Figure S4| Temperature dependent steady state photoluminescence. a,** Evolution of (EDBE)PbCl$_4$ emission profile with temperature (E$_{exc}$ 3.76 eV). The dashed lines indicate the temperature-independent energy of the three principal components (W1=1.96 eV, W2=2.15 eV and W3= 2.34 eV) determined from the principal component fitting of all photoluminescence spectra. **b,** Arrhenius plots of the integrated PL intensity (*I*) as a function of temperature. The refinement was done using the Arrhenius formula $I = I_0/[1 + a\,exp(-E_a/kT)]$, where $E_a$ is the activation energy, $k$ is the Boltzmann constant, $I_0$ is the zero-temperature PL intensity and *a* represents the strength of the quenching process. The plots are relative to (EDBE)PbCl$_4$ (upper panel) and (EDBE)PbBr$_4$ (lower panel)

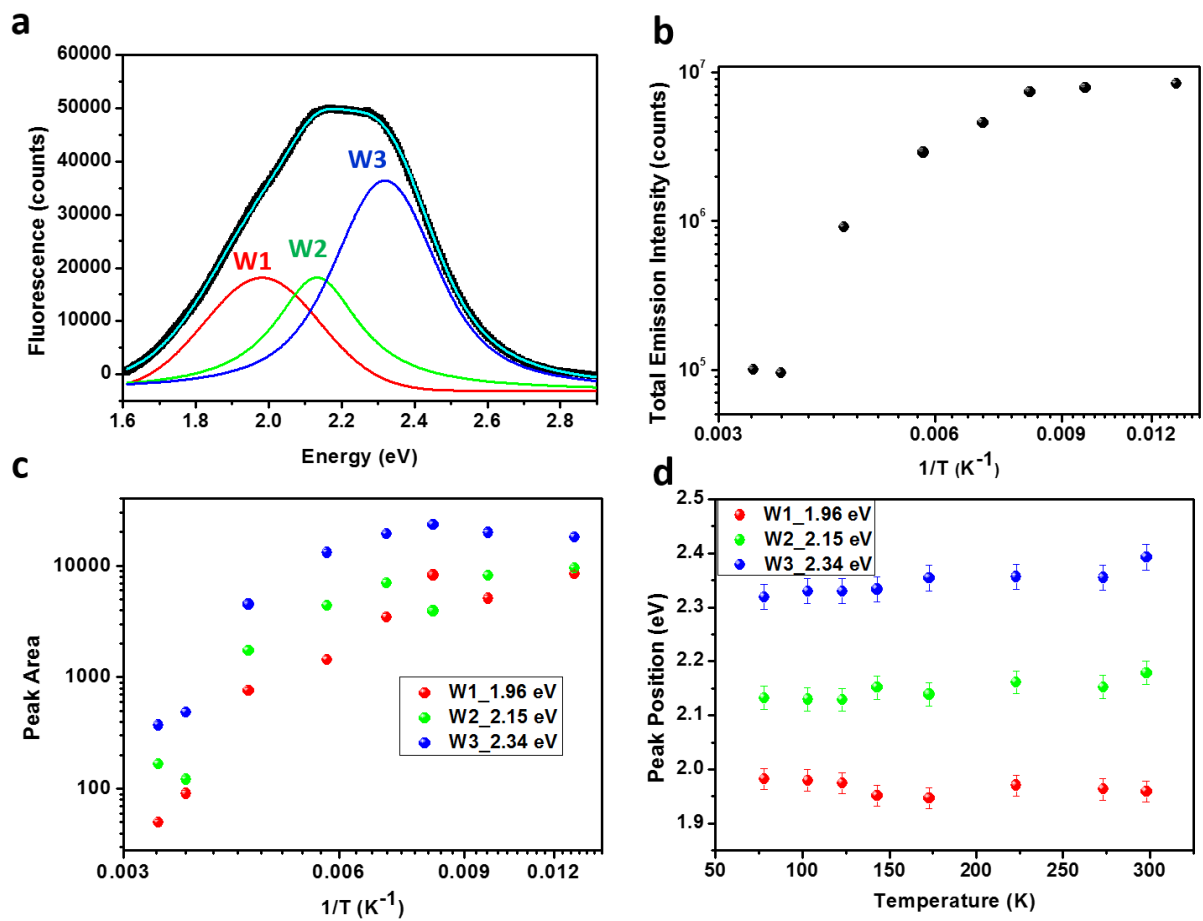

**Figure S5| Temperature dependent steady state luminescence of (EDBE)PbCl$_4$. a,** Steady state PL at 78 K showing the three components W1, W2, W3 peaked at 1.96 eV, 2.15 eV and 2.34 eV, respectively. **b,** Temperature dependence of total emission intensity, showing increase of almost two orders of magnitude at 78 K. **c,** Area of each PL component respect to temperature, showing the different trend of W1, W2 and W3 during the temperature decrease. **d,** Peak position vs temperature, indicating that W1, W2 and W3 do not shift with temperature, but only their relative intensity is affected.

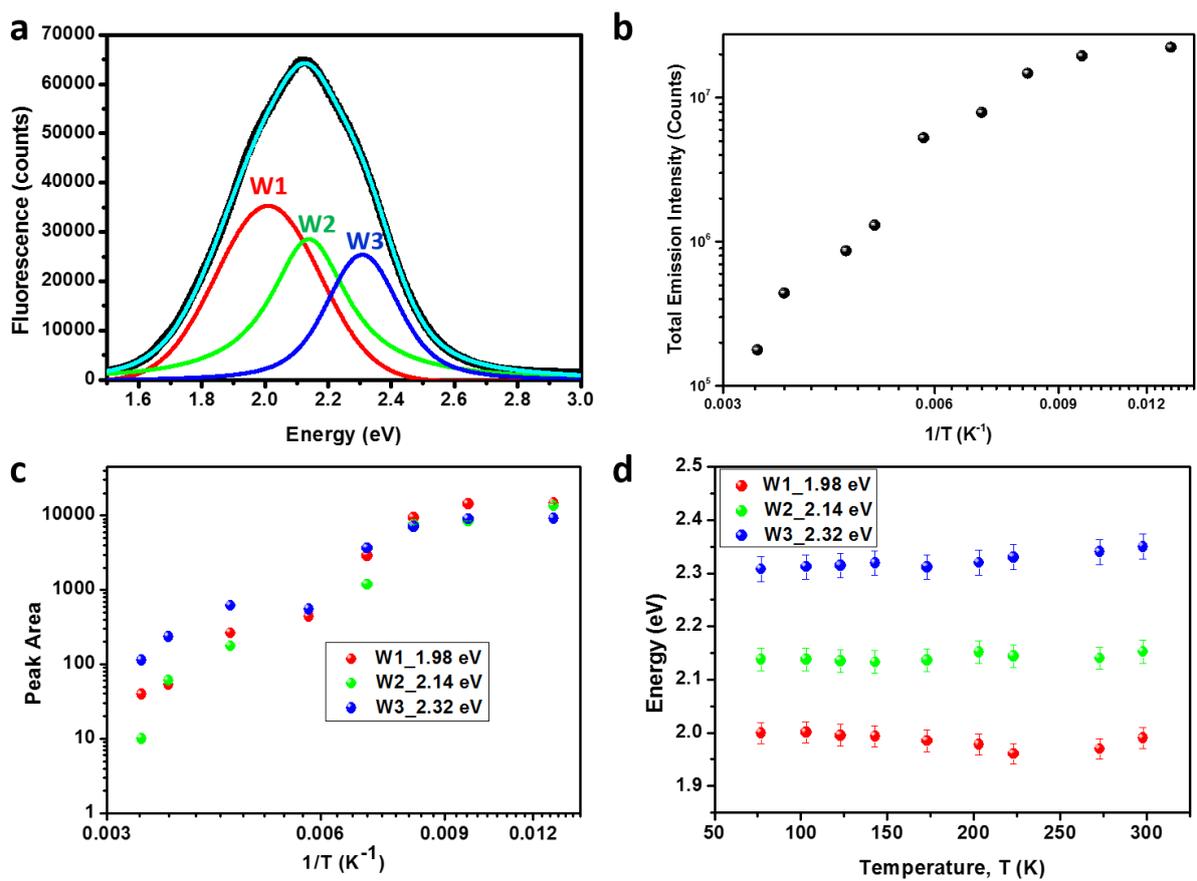

**Figure S6| Temperature dependent steady state luminescence of (EDBE)PbBr$_4$. a,** Steady state PL at 78 K showing the three components W1, W2, W3 peaked at 1.96 eV, 2.15 eV and 2.34 eV, respectively. **b,** Temperature dependence of total emission intensity, showing increase of almost two orders of magnitude at 78 K. **c,** Area of each PL component respect to temperature, showing the different trend of W1, W2 and W3 during the temperature decrease. **d,** Peak position vs temperature, indicating that W1, W2 and W3 do not shift with temperature, but only their relative intensity is affected.

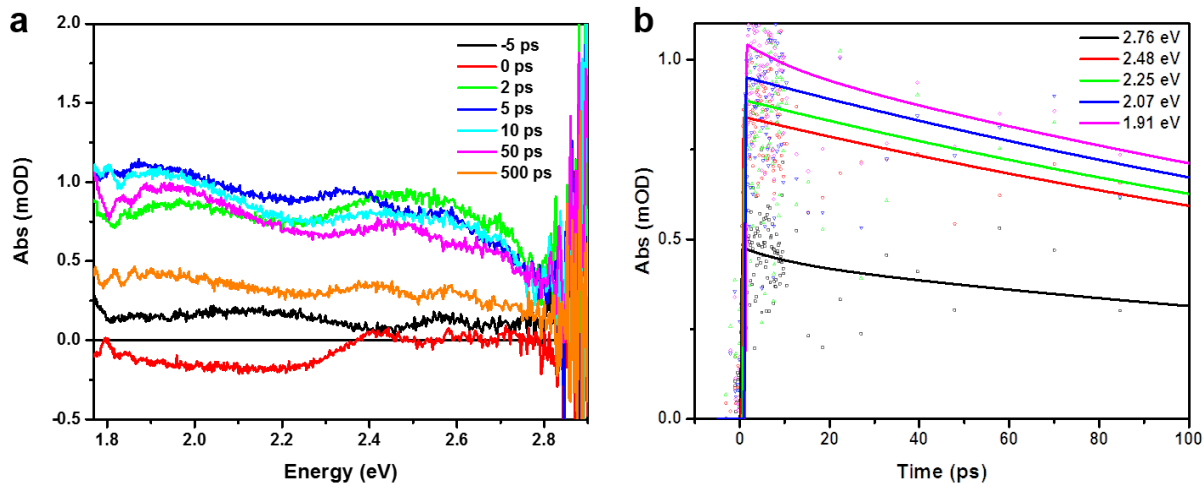

**Figure S7| Transient absorption (TA) measurements of (EDBE)PbCl$_4$. a,** TA spectra with excitation resonant to the excitonic peak (Eexc = 3.76 eV) and **b,** corresponding decay dynamics at different spectral position. The decays follow a similar dynamic across the whole spectral range.

**Table S2| Transient absorption (TA) dynamics fitting parameters for different excitation energy.** The characteristic lifetimes ($\tau$) and amplitudes (A) were extracted from the global fitting of 6 decays at different spectral regions using a double exponential decay function. The excitation energy is resonant to the excitonic peak of the perovskite, $E_{exc}$ = 3.76 eV and $E_{exc}$=3.26 eV for (EDBE)PbCl$_4$ and (EDBE)PbBr$_4$, respectively.

| Material | Excitation Energy (eV) | Probe Energy (eV) | τ1 (ps) | A1 | τ2 (ps) | A2 |
|---|---|---|---|---|---|---|
| (EDBE)PbCl$_4$ | 3.76 | 2.76 | 11.1 | 0.08 | 296.4 | 0.92 |
|  |  | 2.48 |  | 0.07 |  | 0.93 |
|  |  | 2.25 |  | 0.00 |  | 1.00 |
|  |  | 2.07 |  | 0.04 |  | 0.96 |
|  |  | 1.91 |  | 0.05 |  | 0.95 |
| (EDBE)PbBr$_4$ | 3.26 | 2.76 | 9.8 | 0.46 | 741 | 0.54 |
|  |  | 2.48 |  | 0.51 |  | 0.49 |
|  |  | 2.25 |  | 0.46 |  | 0.54 |
|  |  | 2.07 |  | 0.41 |  | 0.59 |
|  |  | 1.91 |  | 0.19 |  | 0.81 |

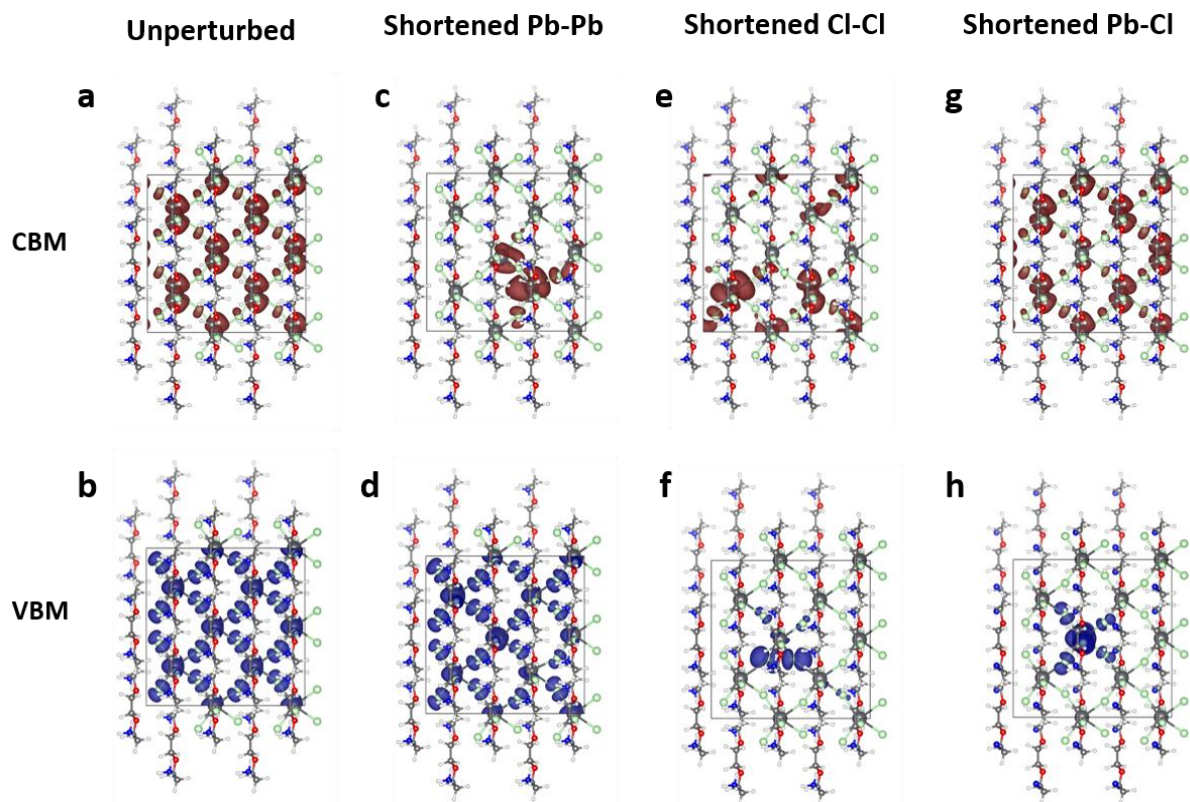

**Figure S8| Plots of charge densities of (EDBE)PbCl$_4$ calculated at the PBE/wSOC level.** The figures represent the conduction band minimum (CBM) and valence band maximum (VBM) distributions in the unperturbed crystal structure (**a, b**), and upon application of a perturbation. These perturbations involve the shortening of the following atomic distances: Pb-Pb (**c, d**); Cl-Cl (**e, f**); Pb-Cl (**g, h**). Strong charge localization is induced selectively by the applied lattice distortion, compared to the fully delocalized unperturbed system.

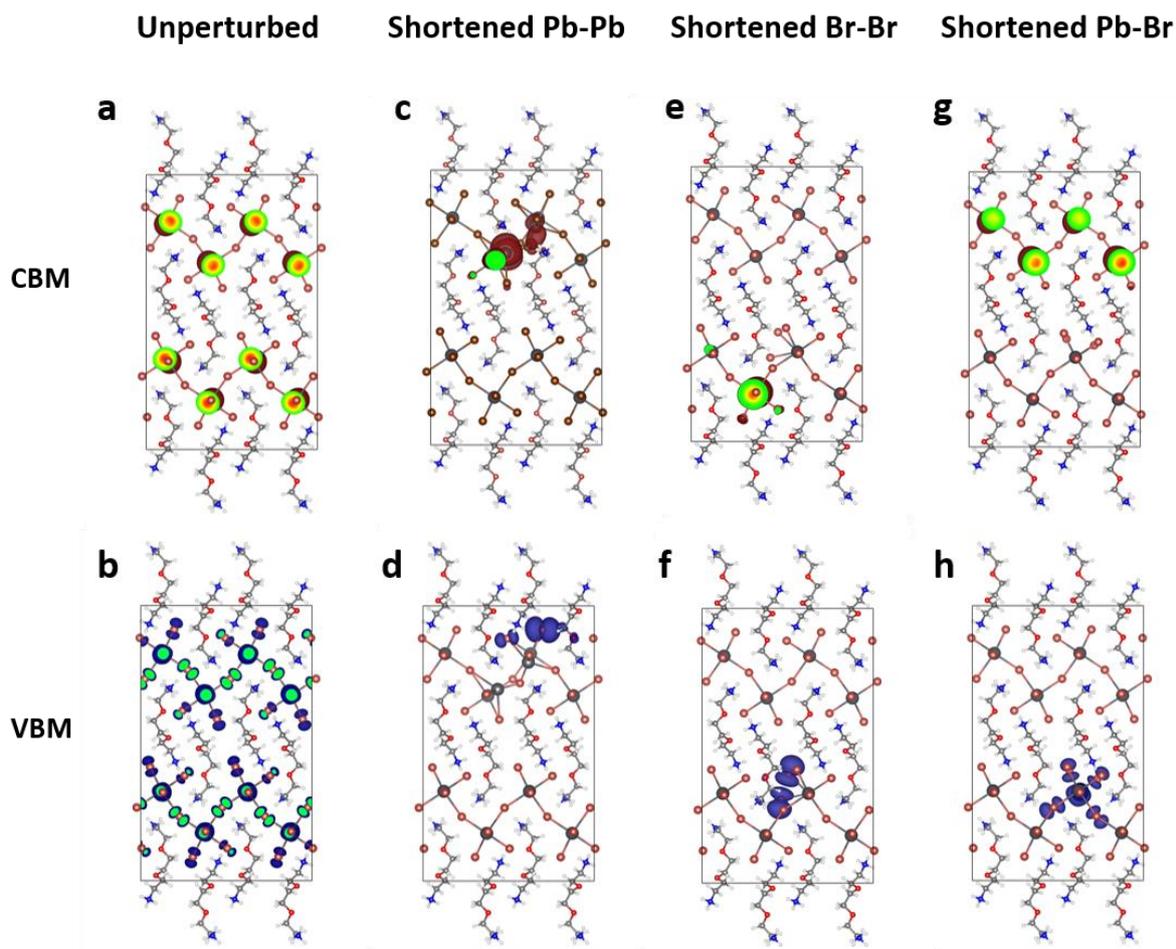

**Figure S9| Plots of charge densities of (EDBE)PbBr$_4$ calculated at the PBE/wSOC level.** The figures represent the conduction band minimum (CBM) and valence band maximum (VBM) distributions in the unperturbed crystal structure (**a, b**), and upon application of a perturbation. These perturbations involve the shortening of the following atomic distances: Pb-Pb (**c, d**); Br-Br (**e, f**); Pb-Br (**g, h**). Strong charge localization is induced selectively by the applied lattice distortion, compared to the fully delocalized unperturbed system.

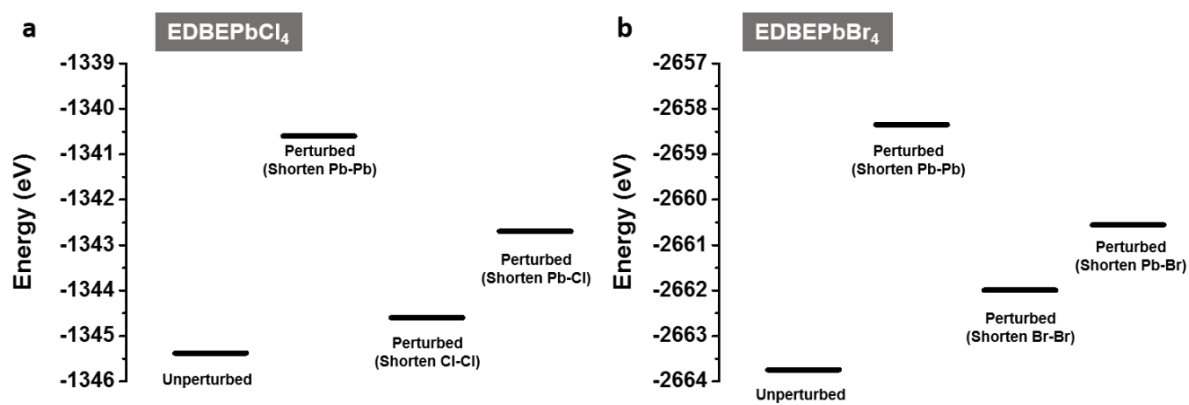

**Figure S10|** Calculated energies at the PBE/wSOC level for the unperturbed and perturbed crystal structures of **a,** (EDBE)PbCl$_4$ and **b,** (EDBE)PbBr$_4$.